# Native Pb vacancy defects induced p-type characteristic in epitaxial monolayer PbSe

Qiwei Tian,[1] Ping Li,[2, a)] Li Zhang,[1] Yuan Tian,[1] Long-Jing Yin,[1] Lijie Zhang,[1, a)] and Zhihui Qin[1, a)]

1. *Key Laboratory for Micro/Nano Optoelectronic Devices of Ministry of Education & Hunan Provincial Key Laboratory of Low-Dimensional Structural Physics and Devices, School of Physics and Electronics, Hunan University, Changsha 410082, China*

2. *State Key Laboratory for Mechanical Behavior of Materials, Center for Spintronics and Quantum Systems, School of Materials Science and Engineering, Xi'an Jiaotong University, Xi'an, Shaanxi 710049, China*

a) Authors to whom correspondence should be addressed:

pli@xjtu.edu.cn; lijiezhang@hnu.edu.cn; and zhqin@hnu.edu.cn

PbSe, a predicted two-dimensional (2D) topological crystalline insulator (TCI) in the monolayer limit, possess excellent thermoelectric and infrared optical properties. Native defects in PbSe take a crucial role for the applications. However, little attention has been paid to the defect induced doping characteristics. Here, we provide an experimental and theoretical investigation of defects induced p-type characteristic on epitaxial monolayer PbSe on Au(111). Scanning tunneling microscopy (STM) measurements demonstrate an epitaxial PbSe monolayer with a fourfold symmetric lattice. Combined scanning tunneling spectroscopy (STS) and density functional theory (DFT) calculations reveal a quasi-particle bandgap of 0.8eV of PbSe. STM results unveil that there are two types of defects on the surface, one is related the vacancies of Pb atoms and the other is the replacement of the absent Se atoms by Pb. Corresponding theoretical optimization confirms the structures of the defects. More importantly, both STS measurements and DFT calculations give evidence that the Pb vacancies move the Fermi energy inside the valence band and produce extra holes, leading to p-type characteristics of PbSe. Our work provides effective information for the future research of device performance based on PbSe films.

Group-IV monochalcogenides have widely been studied for their excellent applications in ferroelectric,[1,2] thermoelectric,[3-5] and photoelectric devices.[6] For example, GeSe has a ferroelectric to antiferroelectric phase transition regulated by electric field.[7] SnTe possesses a stable in-plane ferroelectricity in atomic-thick due to spontaneous polarization.[8] Jiang *et al.* improved the performance of electron and phonon localization-induced high-entropy GeTe-based thermoelectric materials by entropy manipulation.[3] As a representative member of Group-IV monochalcogenides, bulk PbSe has a NaCl-type structure with a narrow bandgap.[9] It has shown excellent performance in near infrared photoelectric appliances.[10] Low-dimensional PbSe exhibits excellent physical properties due to the significant quantum effects of dimension reduction.[11] PbSe quantum dots (QDs) are promising photovoltaic materials due to their solution workability, strong quantum limiting effect and adjustable bandgap. Recently, Liu *et al.* achieved the highest performance for the infrared (IR) solar cells through efficiently passivated PbSe QD solids.[12] In addition, theoretical calculations predict that PbSe monolayer is a two-dimensional (2D) topological crystal insulator (TCI) with Dirac linear boundary states protected by crystal symmetry.[13] Very recently, Shao *et al.* found evidence of the existence of topological boundary states in few-layer PbSe epitaxial grown on $SrTiO_3$ surface.[14] So far, the topological physics of PbSe in the single-layer limit has not been reported yet. To date, monolayer PbSe have been only grown in very few systems based on functional surfaces.[15,16]

On the other hand, defects play a crucial role in the performance of the devices based on Group-IV monochalcogenides. Jiang *et al.* achieved high ZT in GeTe-based thermoelectric materials induced by evolution of defect structures.[17] Wang *et al.* achieved high performance thermoelectric properties through defect control in SnSe.[18] Recently, Duvjir *et al.* directly observed the intrinsic defects in SnSe through a scanning tunneling microscope and explained the p-type characteristics induced by Sn-vacancy under theoretical calculations.[19] The electronic and optical properties of PbSe have been described in the previous literatures from the perspective of theoretical calculation and device transport.[20-24] However, direct characterization of the defects-induced electronic properties tuning in monolayer PbSe is lacking.

In this work, we report the structural and electronic properties of epitaxial PbSe monolayers on Au(111) by using scanning tunneling microscopy/spectroscopy (STM/S) and density functional theory (DFT) calculations. Atomically resolved STM topographic images reveal that the epitaxial PbSe monolayer has a fourfold symmetric square lattice. One-dimensional stripe moiré patterns are formed due to the lattice mismatch between PbSe and Au(111). A quasi-particle bandgap of ~0.8eV has been obtained by STS measurements, which is in consistent with the theoretical calculation value of freestanding PbSe. STM results indicate that there are two types of defects on the surface, i.e., Pb vacancy and replacement of Se vacancies by Pb atoms. Experimental and theoretical investigations confirm the Pb vacancy defects induced p-type doping of PbSe.

The monolayer PbSe on Au(111) was prepared in a home-built ultrahigh vacuum (UHV) chamber equipped with a CreaTec LT-STM system with the base pressure below $1\times10^{-10}$mbar. The clean Au(111) surface was achieved by several cycles of $Ar^+$ sputtering with an ion energy of 1000eV and subsequent annealing up to 700K. We first deposited Pb (99.999%, purchased from *Alfa Aesar*) on the Au(111) substrate maintained at room temperature from a home-built evaporation cell with a flux of ~0.1ML/min. The Pb was evaporated at the temperature of ~800K for ~10min. Subsequently, Se (99.999%, purchased from *Alfa Aesar*) was deposited on the Pb/Au(111) surface at room temperature. The source temperature of Se is 400K from another home-built evaporator in the system. Then the sample was annealed at 500K for 20min and transferred in situ to the STM chamber for measuring under liquid-nitrogen temperature (77K). The STS measurements were acquired by a lock-in technique (739Hz, 30mV AC bias modulation). The bias voltages were respected on the sample, and the topographic images were taken using a constant-current mode. All the STM topography images were processed using WSxM software.[25]

First-principles calculations were performed using Vienna *ab initio* simulation package (VASP)[26] with Perdew Burke Ernzerhof (PBE) functionals[27] in the framework of generalized gradient approximation (GGA). A vacuum of 30Å was used to avoid the interaction between the samples with their images. The self-consistent convergence criteria for electronic structure was $10^{-6}$eV, and the atomic positions are relaxed until the residual force on each atom is less than 0.01eV/Å. The plane wave energy cutoff in the ground state was set to 500eV, and 9×6×1 and 15×12×1 Γ-centered k-mesh samplings were adopted for structural optimization and self-consistent calculations. A four-layer Au(111) slab was used to simulate the substrate. In the simulation, the surface reconstruction of Au slab was ignored. The van der Waals interaction between the PbSe and the Au flake was considered using DFT-D3 dispersion correction.[28] The STM images were calculated through the Tersoff-Hamann method.[29]

The intrinsic PbSe crystal has a NaCl-type structure at ambient pressure,[9] consisting of two interlaced face-centered cubic (fcc) structures [see the ball and stick model shown in Fig. 1(a)]. The (001) surface of PbSe exhibits a fourfold symmetry with a lattice constant of 0.43nm, whereas a unit cell thickness of 0.6nm along the [001] direction. The monolayer PbSe was epitaxially grown on Au(111) by the two-step method in UHV, as shown the schematic diagram in the top panel of Fig. 1(b). We first deposited ~1 ML Pb on Au(111) substrate, observing a large number of Pb clusters distributed on the surface from the STM image [see the bottom panel of Fig. 1(b)]. Figure 1(c) shows a large-scale STM image of the formed PbSe after Se deposition and subsequent annealing. The obtained homogeneous monolayer film growing on the surface accompanied by 1D moiré patterns due to lattice mismatches between PbSe and Au(111) with a period of ~2.15nm [see also in Fig. S1(a)]. The height profile in Fig. S1(b) reveals the monolayer feature of PbSe. High resolution STM image shown in Fig. 1(d) displays a single-layer PbSe highlight by a black box in Fig. 1(c). In addition, there are abundant of point defects decorated on the surface. The atomic resolved STM image zoom-in the defects-free region of Fig. 1(d) demonstrates a fourfold symmetric lattice with a lattice constant of ~0.43nm, which is consistent with previous literature reports.[14,15] The single-layer PbSe film on the surface of Au(111) did not undergo reconstruction but maintain the 1×1 structure. By changing the scanning bias, we found a distinct 1D stripe moiré pattern caused by the lattice mismatch between the tetragonal and the hexagonal lattice materials.[30] The fast Fourier transform (FFT) results shows two sets of spots marked by red and white circles as shown in the inset of Fig. 1(g), in accord with the periodicity of moiré superstructure and atomic lattice, respectively. The inverse FFT-filter STM image created using the peaks in the inset of Fig. 1(g) clearly visualize the periodicity with perfect crystalline. We carry out differential conductance (dI/dV spectrum) on the position marked in Fig. 1(e) far away from the defects. The valence band maximum (VBM) and conductive band minimum (CBM) are determined at ~0.4eV and 0.4eV, respectively, by a highly accepted method,[31] as shown in Fig. S2, revealing a bandgap of 0.8eV. It is worth mention that our experimental results also show intrinsic characteristics. In the previous literature,[14] PbSe films grown on the STO surface showed a bandgap associated with the thickness of the layer, which is familiar to the transition metal dichalcogenides (TMDC) materials. It has a bandgap of 0.4eV at the thickness of six layers and 1.3eV for three layers. The reduced bandgap of monolayer PbSe on the Au(111) is the fact that the electrons derive from precious metal surface are prone to hybridization with the 2D materials, which is also the case in the contacts of the metal electrodes and 2D materials.

In addition, we performed the first-principles calculations based on DFT. First, we notice that the 2×3 unit cell of PbSe is commensurate to the 3×√21 Au(111) with a lattice mismatch of less than 0.3%. The lattice mismatch is defined as $\Delta = 1/a_{Au}|a_{PbSe} - a_{Au}|$, where $a_{Au}$ and $a_{PbSe}$ correspond to the lattice constants of the Au(111) substrate and PbSe, respectively. We have considered three typical configurations, as shown in Fig. 2. After extensive geometry optimizations, we found that the total energy of the three structures is almost identical. Figure 2(d-f) show the DOS of PbSe detached from Au(111), and we found that the DOS of the three structures is exactly the same. On the other hand, theoretical calculations show that PbSe is an intrinsic semiconductor with a bandgap of 0.7eV, which is very close to our experimental findings.

Defects play an important role in tuning the electronic properties of 2D materials and heterostructures.[32-35] We systematically studied the defects in epitaxial PbSe films on Au(111). As shown in Fig. 3, a number of defects exist in the epitaxial grown monolayer PbSe on the Au(111) surface. In the filled state STM image, obvious atom absence (dark) and outstanding bright spots, respectively, are shown, attributed to two types of defects. We defined them as type-A and type-B defects, respectively. We proposed the detailed defect structure as shown in Fig. 3b by considering the previous work of Group-IV monochalcogenides.[19,23,36] We did not find the bias dependent feature of the defects. As shown in the filled state STM image, we propose that it is the absence of Pb atoms on the surface for the type-A defect structure, and the corresponding atomic structure model is established. The simulated STM image results are consistent with the experimental results. For type-B defect structure, the filled state STM image shows five uniform bright spots. As shown in the atomic structure model, the absence of Se atom on the surface is followed by additional Pb atom occupying the site, forming the defect structure in which Pb atom replaces Se atom. The simulated STM image are consistent with the experimental results, confirming the defect structure of the substituted type. In addition, in the epitaxial PbSe monolayers, the type-A defect structure is dominant and has different types. As shown in Fig. 3(a), for the absence of a single Pb atom, the STM image shows three types of hole morphology, namely the circular type marked with (i), the triangular type marked with (ii), and the diamond type marked with (iii). Figure 3(c) shows the atomic structure models of the above-mentioned three defect structures in detail. For the circular type A defect, a single Pb atom is missing. In addition to the absence of Pb atom in the center, the defect of type-A triangle is accompanied by the absence of Se atom in the nearest neighbor and finally shows the triangle shape. As for the diamond-shaped type-A defect form, there is the absence of Pb atom in the center, accompanied by the absence of Se atom in the nearest diagonal, and finally presents the diamond-shaped type-A defect form.

In order to investigate whether the doping characteristics is related to the defects, we carry out STS measurements on different types of the defects and the area nearby. A typical STM image of type-A defect (i) is shown in the inset of Fig. 4(a). We record dI/dV spectra on different positions highlighted by red, blue and black crosses, respectively, marked on the STM image. The top panel shows a black smooth curve of dI/dV spectra on the native Pb vacancy defect. The other two spectra shown in the bottom panel of Fig. 4(a) with multi-small peaks of defect states are highlighted by a black arrow, revealing that the defect is negative charged whereas the surrounding is a donor.[37] It is obvious to deduce that the characteristic from the spectra exhibits a p-type doping because the Fermi level is located closer to the valence band.[38-40] In fact, we confirm that the Pb vacancy acts, actually, as an acceptor responsible for the p-type characteristics of PbSe. Meanwhile, the dI/dV spectrum conducted on type B defects shows an intrinsic semiconductor. To support our finding, we carry out first-principle

DFT calculations to examine electronic properties effects of these two types of defects in PbSe. For TDOS shown in Fig. 4(c) of the type A defects, the vacancy Pb moves the Fermi level significantly inside the valence band and then creates extra holes, indicating that PbSe exhibits p-type character,[19] which is in accord with the experimental observation. Meanwhile, the TDOS of type-B defects, as shown in Fig. 4(d), in line with the experiments with a weak n-type doping, displays an obvious in-gap state. The absent of the in-gap state in dI/dV spectra [see Fig. 4(b)] might be due to the strong interaction with Au(111) substrate effects.[41]

In summary, we report on the controlled growth as well as structural and electronic properties of an epitaxial PbSe monolayer on Au(111) investigated using combined STM/STS and DFT calculations. STM imaging with atomic resolution reveals that the epitaxial PbSe monolayer has a fourfold symmetry surface structure and one-dimensional stripe moiré patterns due to lattice mismatches between PbSe and Au(111). STS measurements reveal a quasi-particle bandgap of 0.8eV, matching the value obtained by theoretical calculation of PbSe. More importantly, STS and DFT calculation results reveal that the vacancy of Pb atoms in PbSe leads to a p-type doping.

See the supplementary material for extra experimental data of this work.

This work was supported by the National Natural Science Foundation of China (Grant Nos. 51972106, 12174096, 12204164, 12004295, and 12174095), the Strategic Priority Research Program of Chinese Academy of Sciences (Grant No.

XDB30000000), and the Natural Science Foundation of Hunan Province, China (Grant No 2021JJ20026). The authors acknowledge the financial support from the Fundamental Research Funds for the Central Universities of China. P. L. thanks China's Postdoctoral Science Foundation funded project (No. 2022M722547).

AUTHOR DECLARATIONS

Conflict of Interest

The authors have no conflicts to disclose.

## Author Contributions

**Qiwei Tian:** Conceptualization (equal); Formal analysis (lead); Investigation (lead); Methodology(lead); Validation (lead); Writing – original draft (lead). **Ping Li:** Methodology (equal); Software(equal); Writing – review & editing (equal). **Li Zhang:** Investigation (supporting). **Yuan Tian:** Investigation (supporting). **Long-Jing Yin:** Investigation (supporting). **Lijie Zhang:** Conceptualization (lead); Data curation (lead); Formal analysis (lead); Funding acquisition (lead); Investigation (lead); Project administration (lead); Resources (lead); Supervision (lead); Writing –original draft (equal); Writing – review & editing (lead). **Zhihui Qin:** Investigation (supporting); Project administration (lead); Supervision (equal).

DATA AVAILABILITY

The data that support the findings of this study are available from the corresponding author upon reasonable request.

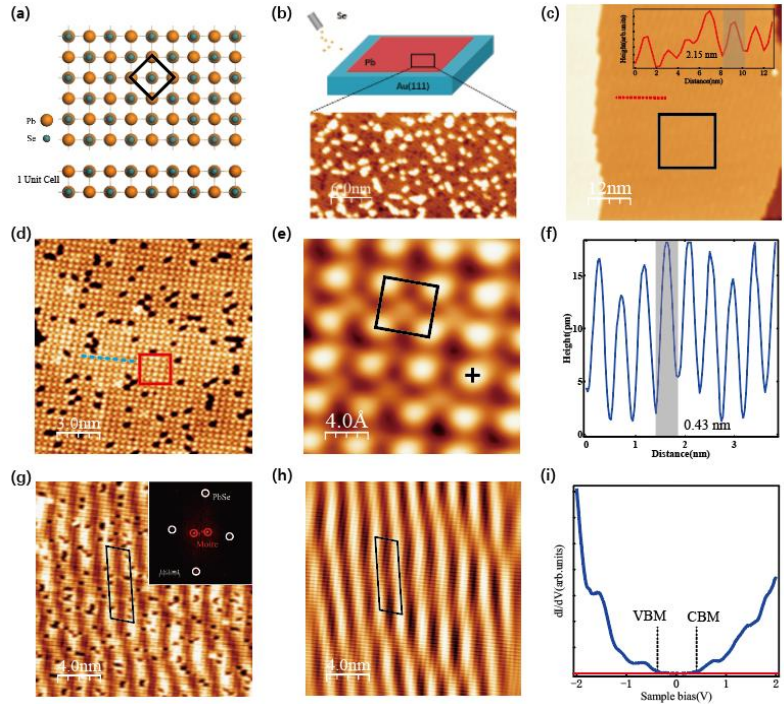

**FIG. 1.** Epitaxial PbSe monolayer on Au(111). (a) Top and side views of ball and stick model of PbSe. (b) A two-step process for growing PbSe. The STM image below shows a precoated 1.2 ML Pb film on the surface of Au(111). (c) The Au(111) is covered with a epitaxial PbSe monolayer. Inset: Line profiles along the red dots line. (d) The enlarged STM image from the area in the black box in (c), showing fourfold symmetry lattice PbSe with a large number of surface defects. (e) Atomically resolved PbSe films are derived from the marked red box in (d). The black wire frame shows the PbSe cell. (f) Line profiles along the blue dot line in (d), revealing the lattice constant of 0.43 nm. (g) STM image at larger sweep bias shows a distinct moiré pattern due to lattice mismatch between Au(111) and PbSe. Inset: The fast Fourier transform (FFT) image of (g) shows two distinct sets of lattice points i.e. the PbSe lattice points of the quadripartite and the moiré pattern of 1D. (h) The inverse FFT image from (g) shows the same striped moiré pattern. (i) Typical dI/dV curve record on PbSe far away from the defects. Scanning conditions are as follows: (b) $V_S =1V$, $I_T=200pA$. (c) $V_S= -2V$，$I_T = 200pA$; (d) $V_S =-10mV$, $I_T=1000pA$; (e) $V_S =-10 mV$, $I_T=1000pA$; (g) $V_S= -100mV$，$I_T = 1000pA$.

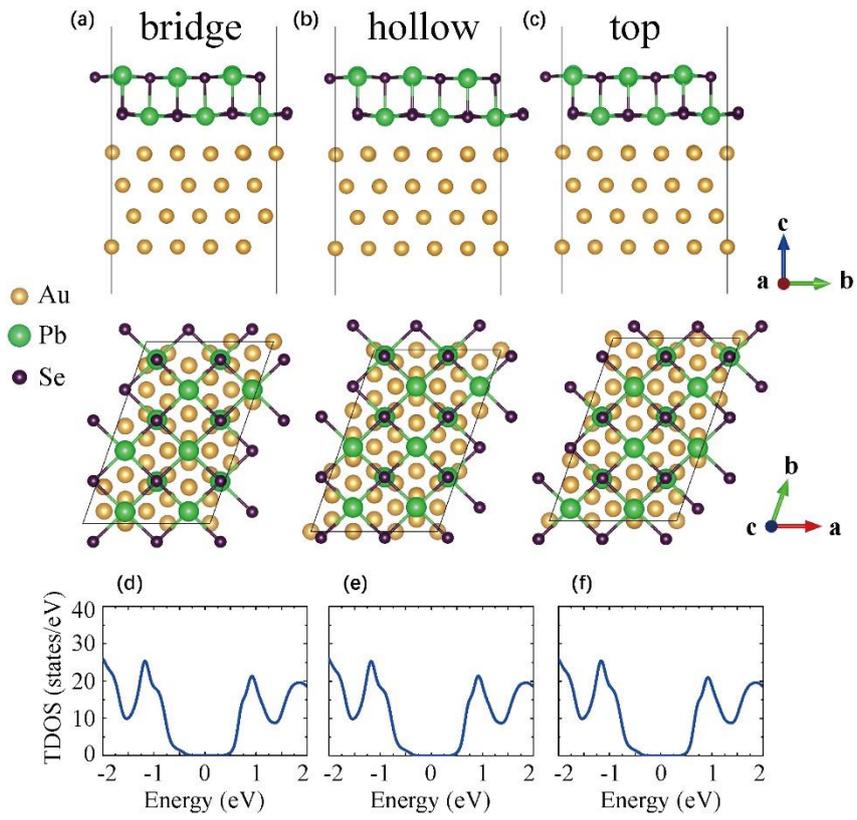

**FIG. 2.** Calculate the DOS of PbSe detached from Au(111) in different stacking configurations. (a)-(c) are stacked on the surface of Au(111) bridge, hollow, and top, respectively. (d)-(f) are the DOS of the corresponding structure.

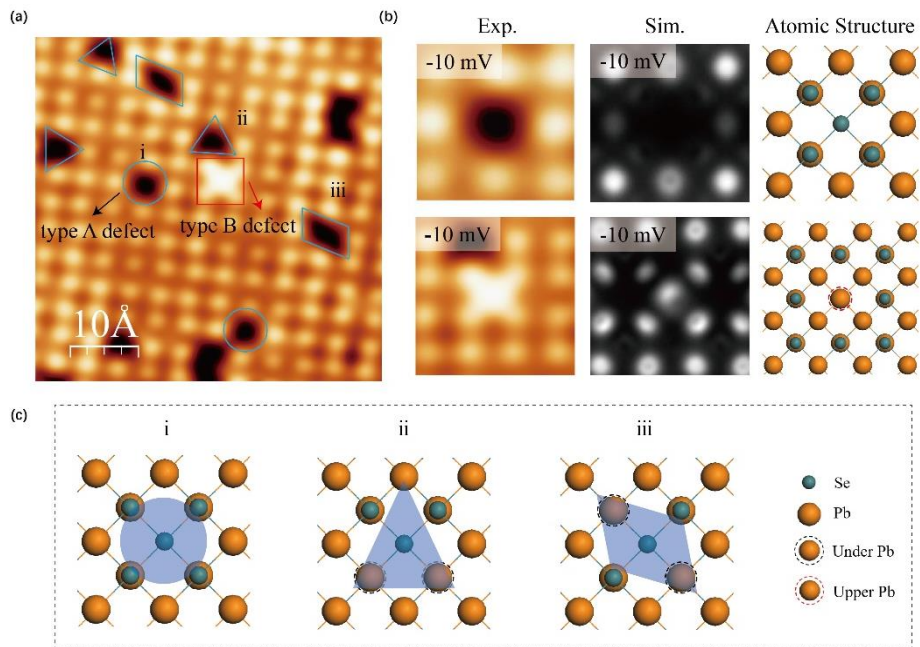

**FIG. 3.** Defect types of PbSe on Au(111). (a) STM image shows two main types of defects in the PbSe surface identified by type-A and type-B defects, respectively. (b) Top panel: Experimental STM image, DFT simulation STM image and atomic structure model of type-A defect. Bottom panel: Experimental STM image, DFT simulation STM image and atomic structure model of type-B defect. (c) Three atomic structure models of type-A defects. The highlight dash areas exhibit a circle, triangle and parallelogram correspond to the area marked with the same shapes in (a). Scanning conditions are as follows: (a) and (b) $V_S = -10$ mV, and $I_T = 1000$ pA.

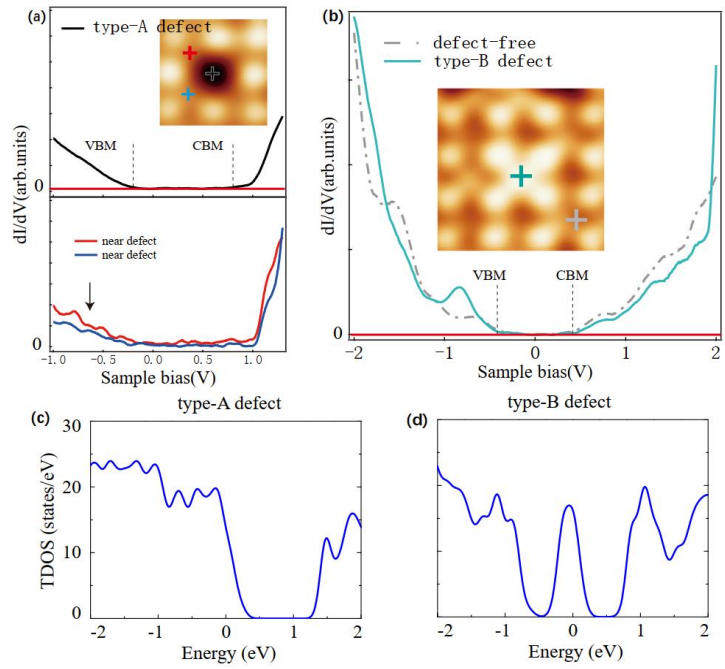

**FIG. 4.** (a) dI/dV spectra recorded on defect and near defect regions. VBM is located much closer to the Fermi level, indicating a p-type semiconductor. Inset: Typical STM image of type-A defect. (b) dI/dV spectra recorded on type-B defect and defect free area, indicating a weak n-type character. (c) Calculated TDOS of type-A defect. (d) Calculated TDOS of type-B defect. Initiate set point: (a-b) $V_S$ =-10mV, $I_T$=1000pA.